\documentclass[conference]{IEEEtran}
\IEEEoverridecommandlockouts
\usepackage{cite}
\usepackage{amsmath,amssymb,amsfonts}
\usepackage{algorithmic}
\usepackage{graphicx}
\usepackage{textcomp}
\usepackage{xcolor}
\usepackage{url}
\usepackage{listings}
\usepackage{subcaption}
\usepackage{balance}
\usepackage{booktabs}

\def\BibTeX{{\rm B\kern-.05em{\sc i\kern-.025em b}\kern-.08em
    T\kern-.1667em\lower.7ex\hbox{E}\kern-.125emX}}
\begin{document}

\title{Energy Efficiency of Web Browsers in the Android  Ecosystem}

\author{
  \IEEEauthorblockN{Nélson Gonçalves\IEEEauthorrefmark{1}\IEEEauthorrefmark{3},
    Rui Rua\IEEEauthorrefmark{1}\IEEEauthorrefmark{3}
Jácome Cunha\IEEEauthorrefmark{2}\IEEEauthorrefmark{3},
Rui Pereira\IEEEauthorrefmark{1},
João Saraiva\IEEEauthorrefmark{1}\IEEEauthorrefmark{3}}

\IEEEauthorblockA{\IEEEauthorrefmark{1}Department of Informatics, University of Minho\\}
\IEEEauthorblockA{\IEEEauthorrefmark{2}Faculty of Engineering of University of Porto\\
\IEEEauthorblockA{\IEEEauthorrefmark{3}HASLab/INESC TEC\\}}
\IEEEauthorblockA{a76268@alunos.uminho.pt, rui.a.rua@inesctec.pt , jacome@fe.up.pt, ruipereira@di.uminho.pt, saraiva@di.uminho.pt}
}

\maketitle

\begin{abstract}

This paper presents an empirical study regarding the energy consumption of the most used web browsers on the Android ecosystem. In order to properly compare the web browsers in terms of energy consumption, we defined a set of typical usage scenarios to be replicated in the different browsers, executed in the same testing environment and conditions. The results of our study show that there are significant differences in terms of energy consumption among the considered browsers. Furthermore, we conclude that some browsers are energy efficient in several user actions, but energy greedy in other ones, allowing us to conclude that no browser is universally more efficient for all usage scenarios.

\end{abstract}


\section{Introduction}
\label{sec:intro}

The use of non-wired, portable devices is changing our
lives. Everyday tasks, such as reading a newspaper, socializing with
friends, listening to music/videos, or the search for information are now
easily performed via an always present mobile phone. Most of those
tasks rely on the internet, and web browsers are one of the most
important and used tools to access the web
\cite{10.1145/2872427.2883014,10.1145/3366423.3380292,
  10.1145/3361242.3361249, 10.1145/3341617.3326142}. The widely use of
mobile devices are also changing the way software engineers develop
their programs: while in the last century runtime performance was
the main goal, in our mobile-device age energy consumption is a new
concern for application software
developers~\cite{pinto2017energy}. Being a web browser one of the most
used software applications it is no surprise the large number of
available browsers (currently there are more than $100$ only for
Android). In a mobile ecosystem, where energy consumption is of
concern to the user, it is very relevant to understand which browser
is the most energy-efficient. In fact, depending on the type of
the task being performed, the browser choice may vary.

In this paper, we present our empirical procedure and preliminary study on five
of the most popular/used web browsers available in the Android
ecosystem. Our execution procedure uses Qualcomm's Trepn Profiler~\cite{trepn}, which is suited for Android devices that use Snapdragon
chipsets and uses \textit{RERAN}~\cite{reran} in order to define simulation scripts that aim to emulate real-user interaction with a physical Android device
and browser under test. For each browser, we developed several scripts with RERAN that aim to emulate three different usage scenarios: \textit{i)} video streaming (on YouTube\footnote{www.youtube.com} and Vimeo\footnote{www.vimeo.com}), \textit{ii)} search engine usage and \textit{iii)} social media interaction. By designing this empirical procedure and executing it on a real, physical device, while monitoring its energy consumption, we intend to answer the following research questions:

\begin{itemize}

\item \textbf{RQ1}: \textit{Which mobile browser is the most
  energy-efficient for browsing YouTube?} According to
  Statista~\cite{statistayoutube}, YouTube is the
  most popular online video platform. Understanding which
  browser is the most energy-efficient can help heavily reduce the
  energy consumed during this typical web browsing.

\item \textbf{RQ2}: \textit{Which mobile browser is the most
  energy-efficient for browsing Vimeo?}  Vimeo is an
  alternative to YouTube which has been gaining popularity, with 7.9 million downloads of the mobile application in 2021 alone \cite{statistavimeo}, by providing identical features with a more pleasant user experience. 

\item \textbf{RQ3}: \textit{Which mobile browser is the most
  energy-efficient for performing searches in search engines?} Providing access to search engines is one of the main reasons to use a web browser. For this scenario, we designed usage scenarios for performing keyword searching on Google.

\item \textbf{RQ4}: \textit{Which mobile browser is the most
 energy-efficient for browsing Facebook?} Social networks are among the most used applications and websites on all platforms, with Facebook being the most widely used \cite{statistaface}. Discovering which mobile browser is the most battery-friendly can help reduce energy consumption and can advise users to choose an energy-efficient browser.

\item \textbf{RQ5}: \textit{Which mobile browser is the most energy-efficient overall?} While there may be browsers more suited
to perform specific tasks, understanding overall which is the most energy-efficient one, can further help users select the most adequate browser for a uniformly more energy-friendly experience.
\end{itemize}

To answer these questions we present an empirical study on the energy efficiency of
the most popular browsers available for Android. The results of our study show interesting findings, such as there is no overall winner in all user action scenarios considered. In fact, some browsers are very energy efficient in some usage scenarios, but energy greedy in other ones.

This remainder of this paper is organized as follows. In Section~\ref{sec:related} we present the related work, followed by Section~\ref{sec:met}, which details the methodology followed to select the corpus of web browsers, as well as the process to execute them with realistic user actions and to estimate its energy consumption. Section~\ref{sec:results} contains the analysis and discussion of our preliminary results. In Section~\ref{sec:validity}, we present the threats to the validity of our work and finally we present our
conclusions (Section~\ref{sec:conc}).

\section{Related Work}
\label{sec:related}


Studies aiming to compare the performance of widely used pieces of software have been increasingly appearing in the scientific community~\cite{Rua2020,importanttool,importanttool2,securityper,luis_testing_frames}. Web browsers are not an exception, despite the majority of the efforts being focused on their runtime performance~\cite{importanttool,importanttool2,securityper}. Nevertheless, in terms of the energy efficiency of browsers, there are also some studies analyzing the energy consumption of different tasks within the same browser.

In~\cite{whokill} an infrastructure is presented to measure the
precise energy consumption used by a mobile browser to load and render
websites as well as the energy needed to render individual web
elements, such as cascade style sheets (CSS), JavaScript, images, and
plug-in objects. This was performed in different types of areas like
finance, e-commerce, email, blogging, news, and social networking
sites, at popular websites such as Gmail, Amazon, and many
others. This study shows that downloading and parsing cascade style
sheets and JavaScript consume a significant portion of the total
energy required to render a page for popular sites, and rendering JPEG
images are considerably cheaper than the other formats. It was also
possible to provide concrete recommendations on how to design web
pages without affecting the user experience to reduce the
amount of energy needed to render the page. Thus, the goal of this work is different from ours since we evaluate several categories of everyday tasks.

Hsu et al.~\cite{greenpage} studied how page loading can affect the
the energy efficiency of three widely used Android web browsers, namely
Chrome, Firefox, and Opera. The paper compares the page loading of 150
random websites considering two different internet frequencies, with the purpose of verifying whether there is a greener page loading browser.  Although
this paper shares our goal of comparing web browsers, it only
focus on loading websites and not exploiting their regular use.

A recent approach~\cite{powdroid} developed a simple
command-line tool, \textit{PowDroid}, to measure the energy consumed
by any application running without requiring access to the source
code. It can be used to perform benchmarks and analyze which components are draining more battery.
Three test scenarios were made to prove its efficiency: web browser,
camera, and weather applications. The web browsers Firefox,
Chrome, Edge, Opera, Samsung Browser, and Brave were
tested and the test consisted of searching for a keyword and opening
three websites. Each test took 4 minutes, with the battery level at
50\% at the beginning, screen brightness at 50\%, and audio at 50\%. Their
experiments found that Brave is the most energy-efficient, while
Firefox was the worst with a 33,8\% increase in energy
consumption. The authors justify these results due to the default
built-in ad-blocker in Brave. Our results also show Brave as one of the
greenest browsers, however, there are user tasks such as playing
YouTube videos where it has poor energy efficiency.

We have conducted a preliminary study comparing the energy performance of desktop-based browsers, where we considered Chrome and Firefox~\cite{sustainse2020}. In that study, we used RAPL~\cite{rapl,raplweb} to monitor the energy consumption during the
execution of each script under the same specific test conditions. These scripts were generated with Selenium~\cite{selenium}, where multiple actions were performed to force browsers to simulate user actions, such as
watching videos, scrolling down, etc. This research was a test-bed and preliminary study to serve with advancements and inspiration to
explore this further, observing the results obtained.

\section{Methodology}
\label{sec:met}

The astonishing adoption of smartphones leads to the increasing use of
mobile apps, including mobile browsers. When selecting their
browser, users consider many factors, such as speed, features, resource
management, themes, compatibility, etc. Due to the lack of
information, users do not currently consider the energy efficiency of
the browsers. While many might assume speed directly equates to energy
efficiency, several research works are showing this is, in fact, not
always
direct~\cite{abdulsalam2015using,marcosblp2014,pereira2020spelling}.

This section describes the design of our benchmark to assess the
energy efficiency of web browsers. We start by specifying the
methodology followed to select the browsers considered in our empirical setup. Next, we define the
techniques and tools used to simulate real user interactions in the
selected browsers. Finally, we detail how we executed and measured the
energy consumption of the actions performed over each web browser.

\subsection{Browser Selection}

Nowadays, there are more than $100$ mobile browsers. Considering all the corpus of available browsers in our empirical setup would be extremely time-consuming, since we intended to simulate real user interaction, by manually performing real user inputs, recording them and replicating them with automatic procedures. Furthermore, we needed to gather a limited set of web browsers, containing browsers with significant and representative usage by Android users, by defining specific selection criteria. 

The first selection criteria consisted in considering the top 5 most downloaded web browsers for mobile devices \cite{top5}. In our reference list, Safari was the second browser most used worldwide, but it was excluded from our corpus since it is not available for Android devices. Furthermore, we considered Google Chrome, Samsung Explorer, UC Browser, Opera Mini, and Mozilla Firefox. Information regarding the number of downloads of each selected browser can be found in Table~\ref{tab:browsers5}.

\begin{table}[!tbh]
 \centering
\caption{Number of downloads for the initial set of considered web browsers.}
\begin{tabular}{lc}
\toprule
\textbf{Browsers} & \textbf{Number of Downloads} \\ \midrule
Chrome            & 5mM+                    \\ 
Samsung Explorer   & 1mM+                    \\ 
UC Browser        & 500M+                   \\ 
Opera Mini        & 500M+                   \\ 
Mozilla           & 100M+                   \\ 
\bottomrule
\end{tabular}
\label{tab:browsers5}
\end{table}

Besides considering the 5 most downloaded web browsers, we also intended to select 2 more browsers with significant usage that could compete with the widely used browsers already considered. After a manual inspection of the list of the top 10 web browsers with the most downloads in the Google Play Store, we selected 2 more browsers, Brave and DuckDuck, due to having interesting features and descriptions on their page. Brave states to
offer a factor of 2 to 4 of speed boost on Android, saving data and battery. They also claim that its users can count up to 2.5 extra hours
of browsing time per battery charge. DuckDuck was selected because it claims it provides an additional level of security and privacy. Furthermore, we wanted to discover if providing such features imply an energy overhead.

\subsection{Execution Procedure}

Our study intends to simulate real-user interaction with a browser, using an automatic execution approach. Obtaining actual usage data from users might raise privacy issues that make such an approach avoidable. Replicating the interaction process manually enough times to obtain reliable estimates of energy consumption and reduce the impact of external interference in the monitoring process is also a time-consuming and cumbersome task. Additionally, performing interactions that perform all sorts of tasks possible in a web browser is impossible. Consequently, the RERAN~\cite{reran} tool was used to record a set of actions that intend to represent the execution of some tasks typically performed in web browsers.

RERAN is a record-and-replay tool that allows to record inputs made on the screen of a device (such as touches, gestures) or other input devices and replicate them with or without delay on a device with the same screen dimensions. In this way, we used this tool to manually record the interactions that intend to simulate real-user interaction in each browser and be able to replicate them several times. Although each recorded test is specific for each browser (i.e. a test was defined for each browser), the steps in each test are equivalent. Also, we performed some of the tests several times in order to assure that the tests have negligible differences in terms of runtime and assure that the comparisons to be made are fair.

In order to answer \textbf{RQ1}, three similar scripts were developed. These scripts
emulate the process of browsing \url{www.youtube.com}, searching for one
specific video by typing in the search bar and playing it in 720p. The
videos chosen were the video-clip \textit{Despacito} from \textit{Luis Fonsi}, the short cartoon \textit{Masha and the Bear-Recipe for Disaster} from \textit{Get Movies} and a video-clip of \textit{Paradise}, from \textit{Coldplay}. The first two videos have been selected from the list of most viewed videos on Youtube~\cite{topyt}, while the latter was randomly selected from a list of music videos. This process was not recorded on Opera Mini and DuckDuck browsers because they did not allow the video to be displayed on full screen, leaving two black bars on the screen. Since this did not happen for other browsers and it could influence the results, we decided to not include these two browsers in this experiment.

The strategy to answer \textbf{RQ2} was similar to the one followed for answering \textbf{RQ1}, but only the video
\textit{Paradise} was the same. The other videos were \textit{A
  Mind Sang} and \textit{Shadows In The Sky}, and they were chosen
randomly. Contrary to what happened with Youtube, where two browsers
did not allow putting the video on full screen, on Vimeo, only Firefox and UC allow putting it on full screen. Therefore, these two browsers were not considered in this evaluation.

The strategy used to answer \textbf{RQ3} consisted of three scripts. They
consist of browsing to \url{www.google.com} and using the search bar to research two questions and one recipe. One of the
scripts consists in searching the question \textit{``who was the first man on the moon’’} and then clicking on the NASA website~\cite{nasa_web}, staying on this website for 30 seconds. Afterward, it returns to Google and clicks on the next result (Apolo 11 on Wikipedia~\cite{apollo11}), staying on such a website for 60 seconds (30 seconds on the main text, and scrolling down through details for another 30 seconds). 
The second script searches for \textit{``world war two’’} and clicks on the first entry, pointing to the Wikipedia website~\cite{ww2}. Afterward, it waits 30 seconds while the first paragraph is on the screen, scrolling down the following text for 50 seconds. The last developed script searches ``recipes with chicken’’ on \url{www.google.com} and proceeds to the Delish website~\cite{delish}. Then, it selects the second recipe and simulates the process of reading ingredients and the five steps for the recipe, spending 20 seconds on each. Afterwards, it scrolls up to watch the recipe video and plays the video until the end. This script was replicated for all browsers except DuckDuck, due to its privacy policy, which always erases the browser history and cookies. For that reason, it did not allow the development of a script with the RERAN tool because it was impossible to guarantee that the clicks were in the same place throughout the different executions.

To answer \textbf{RQ4}, the script starts to browse \url{www.facebook.com} where an account was already logged in. Afterwards, the search bar was used to search for \textit{``Jornal de Noticías’’}~\cite{jn} (a Portuguese newspaper) by clicking on history search. After that, it follows to the photos section, where 15 photos of the front page of a newspaper are loaded one by one. It goes back to
the landing page of \textit{``Jornal de Notícias’’} and scrolls down through the feed where news is posted. After that, it goes to the account's saved content, where a video is selected and played for 60 seconds. This test was not conducted on Opera Mini and DuckDuck browsers for the same reason stated when describing the \textbf{RQ1} strategy.

Finally, by combining all the results obtained with the procedures designed to answer the first 4 research questions, we answer \textbf{RQ5}.

\subsection{Script Execution and Energy Measurement}

In order to profile the execution procedure to estimate energy consumption, we used Trepn Profiler~\cite{trepn}. This profiler has been used in several research efforts~\cite{differenttools3, Rua2020}  to obtain energy measurements, providing accurate results in physical devices with Snapdragon chipsets, since it extracts power measurements from its PMIC (Power Management Integrated Circuit). Trepn can be used to profile hardware and resource usage, as well as to obtain power consumption at the system or application level, with a sampling rate of 100 milliseconds.

Different precautions were taken to have consistent data and minimize external interferences to the monitoring process. 
All measurements were performed in the same LG Nexus 5 device running Android 6.0.1, for which Trepn Profiler is known to retrieve accurate values. The brightness level was set to the minimum, and all applications were turned off, only with Trepn service running in the background. We also performed efforts to reduce effects from cold starts, warm-ups, and cache effects. Every script was executed five times, and the script from Facebook was performed in the middle of the night to ensure that no  news was added and the feed was the same for every browser.


%
%

\section{Results and Analysis}
\label{sec:results}

This section presents and analyzes the results obtained through our execution procedure replicated for each of the $7$ considered browsers. Furthermore, in this section, we use statistical methods and graphical information to describe and analyze the results obtained by combining our executing scripts with Trepn Profiler.

\subsection{Results}

\begin{figure*}[!h]
\begin{subfigure}{0.5\textwidth}
\includegraphics[width=\textwidth]{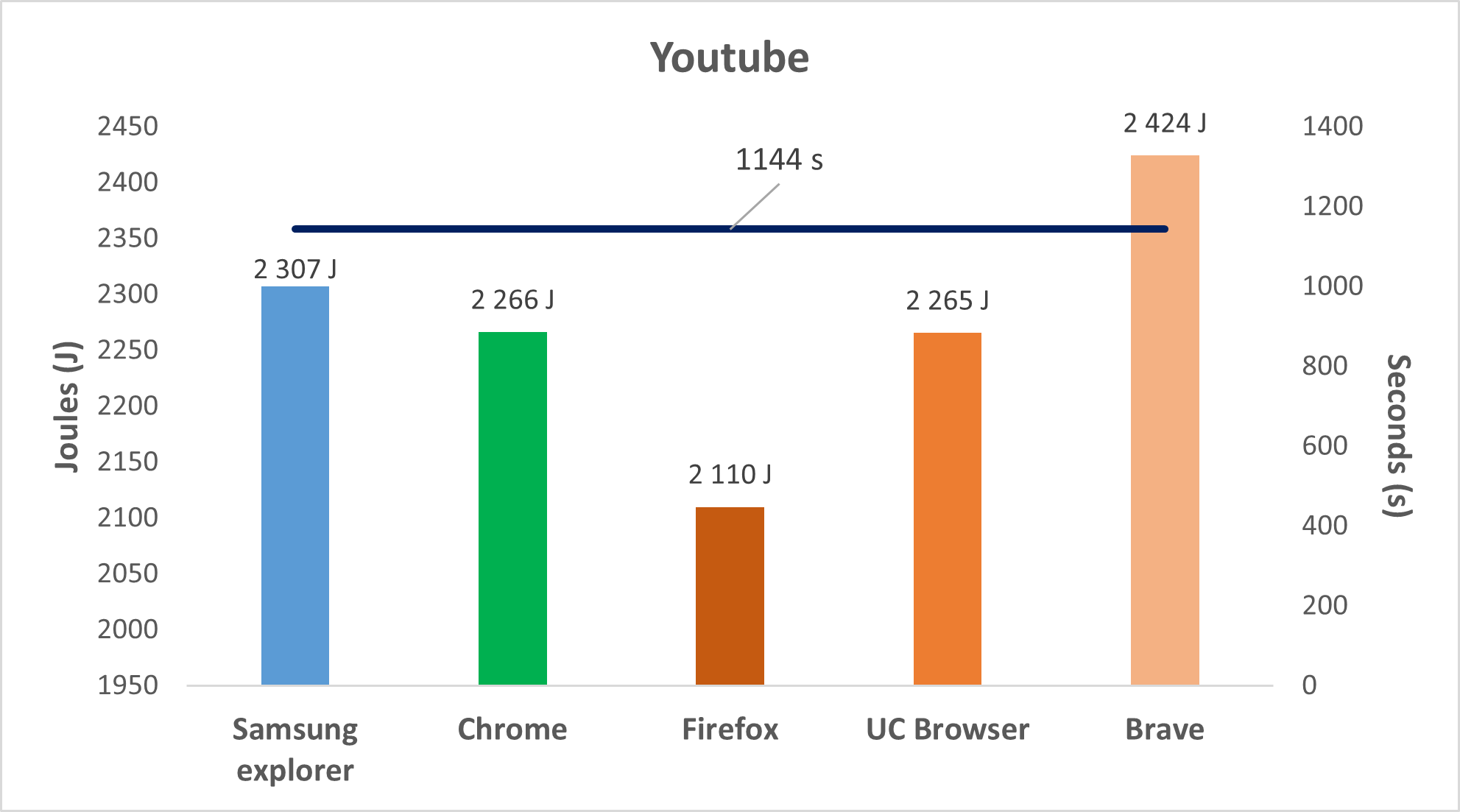}
\caption{Youtube}
\label{fig:Youtube}
\end{subfigure}
\begin{subfigure}{0.5\textwidth}
\includegraphics[width=\textwidth]{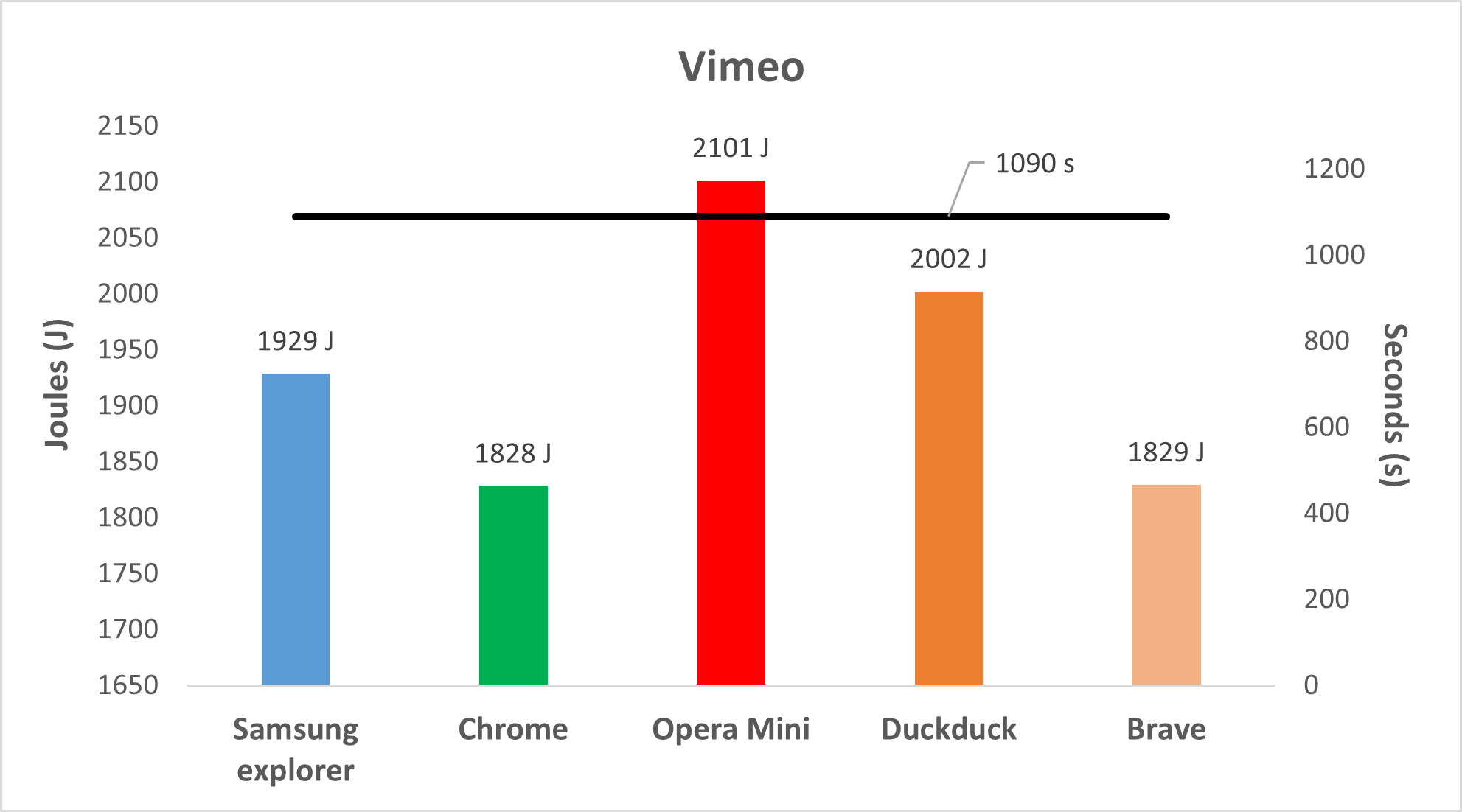}
\caption{Vimeo}
\label{fig:vimeo}
\end{subfigure}
\begin{subfigure}{0.5\textwidth}
\includegraphics[width=\textwidth]{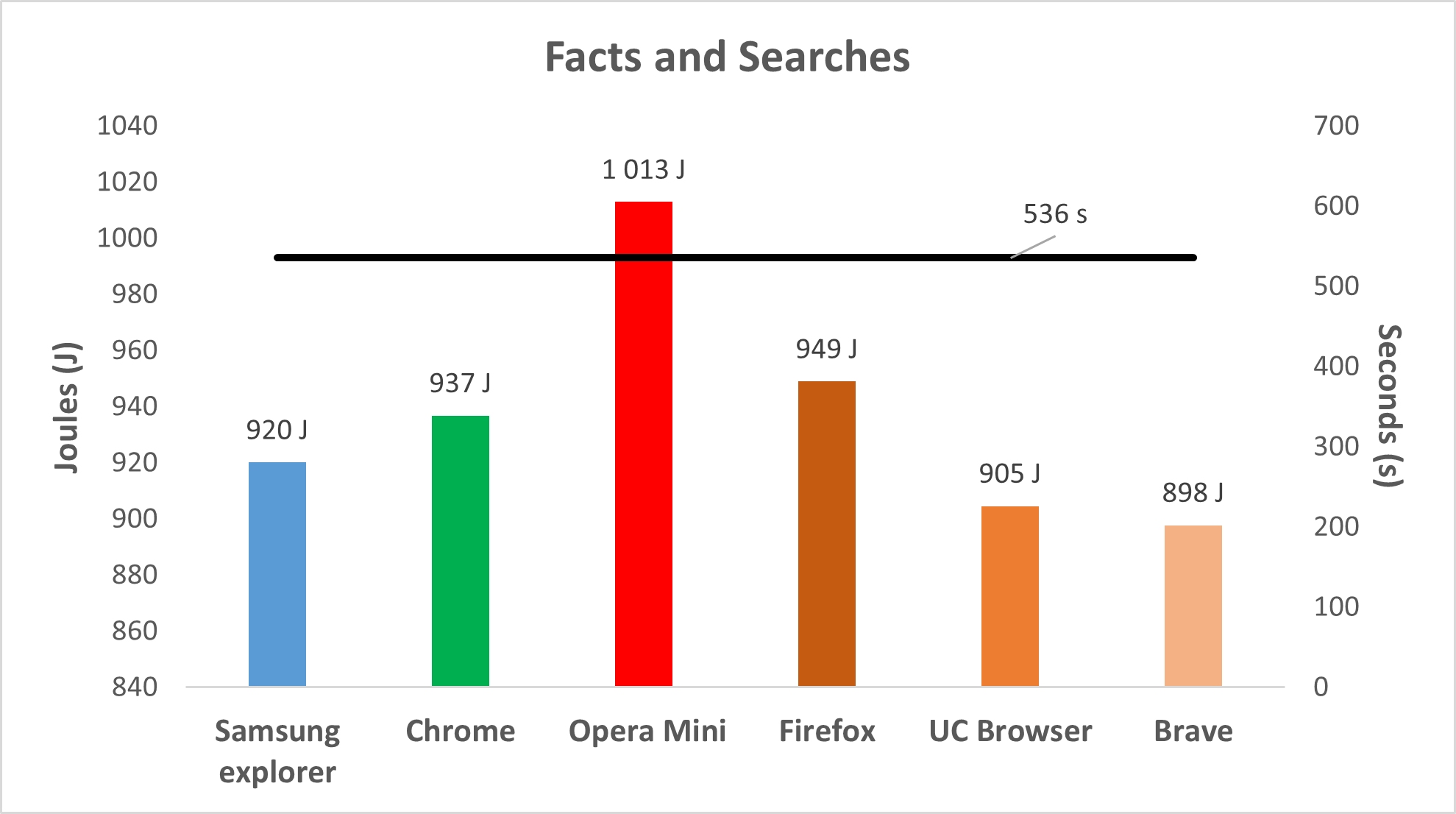}
\caption{Search Engine}
\label{fig:facts}
\end{subfigure}
\begin{subfigure}{0.5\textwidth}
\includegraphics[width=\textwidth]{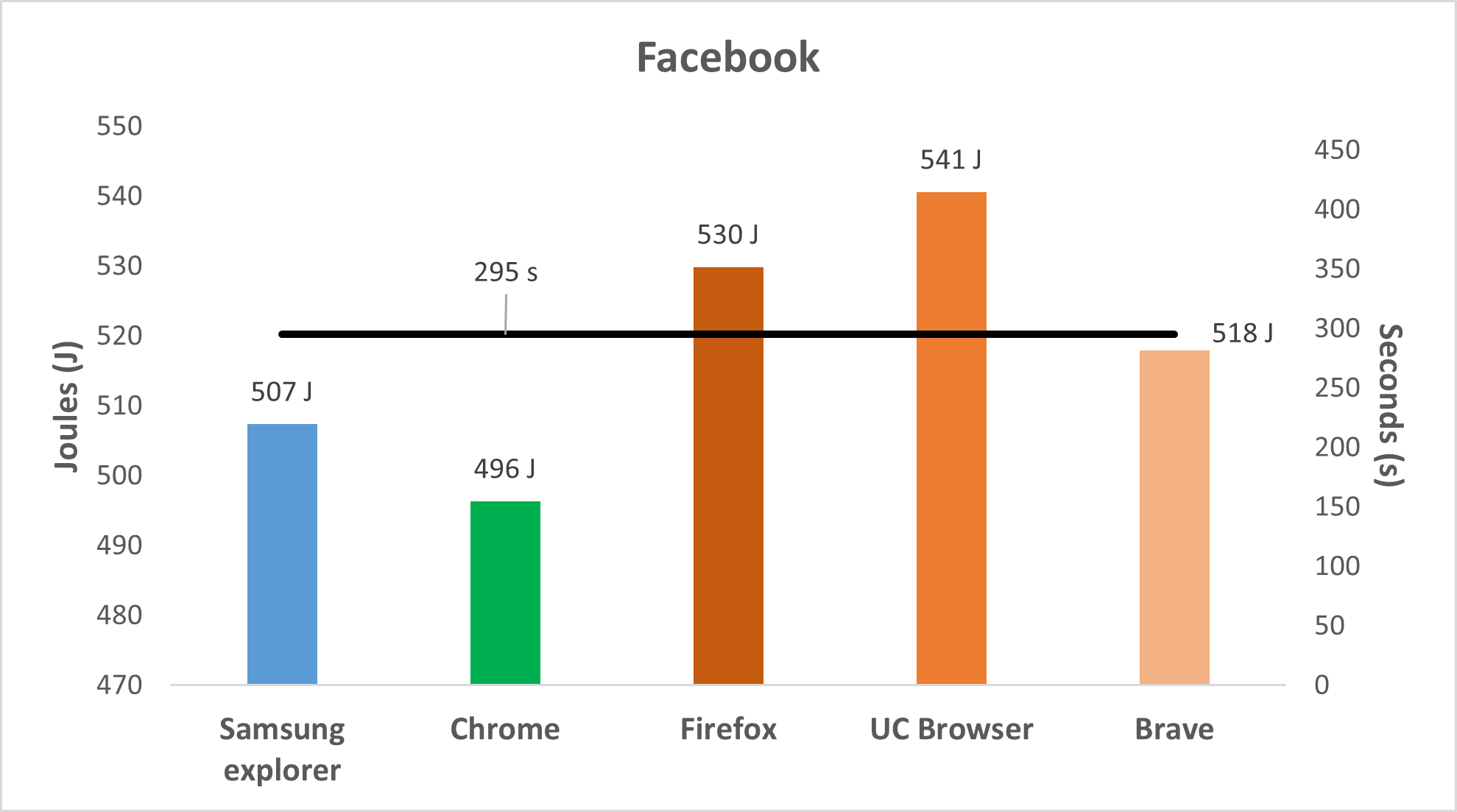}
\caption{Social Media}
\label{fig:facebook}
\end{subfigure}
\caption{Total energy consumption in Joules considering the 3 test.}
\label{fig:energy}
\end{figure*}

To better understand the results obtained by our tests, we display the obtained results in different graphical formats, that present the energy consumed by browsers. Figure~\ref{fig:energy} shows the total energy consumption in Joules
of each browser, for each scenario considered. Each browser has one bar
representing the average of the energy values obtained when executing 5 times each of the 3 developed scripts for each task.
We can also observe a black horizontal line in each of the plots presented in Figure~\ref{fig:energy}, which represents the elapsed time
of the script's execution in each test scenario.

\begin{figure*}[!h]
\begin{subfigure}{0.5\textwidth}
\includegraphics[width=\textwidth]{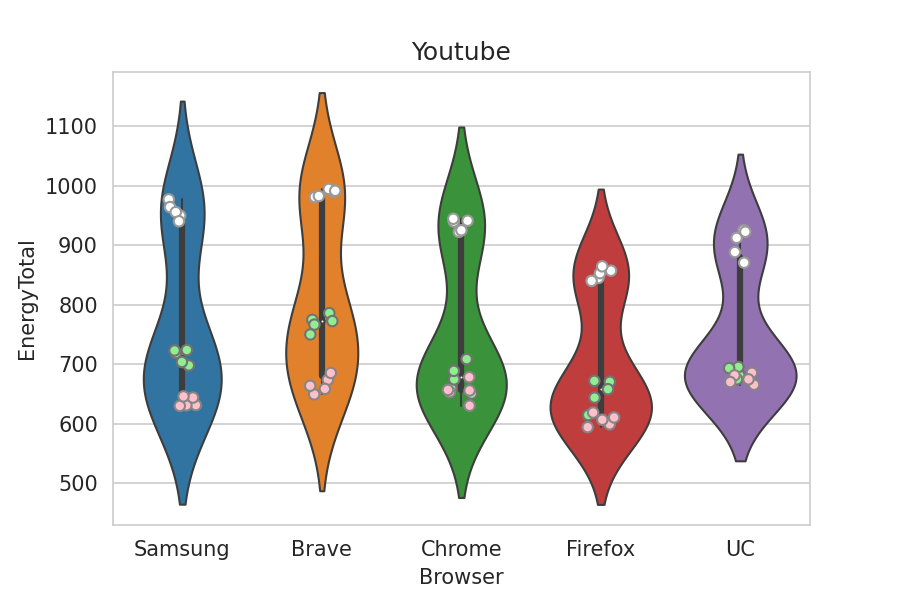}
\caption{Youtube}
\label{fig:Youtube2}
\end{subfigure}
\begin{subfigure}{0.5\textwidth}
\includegraphics[width=\textwidth]{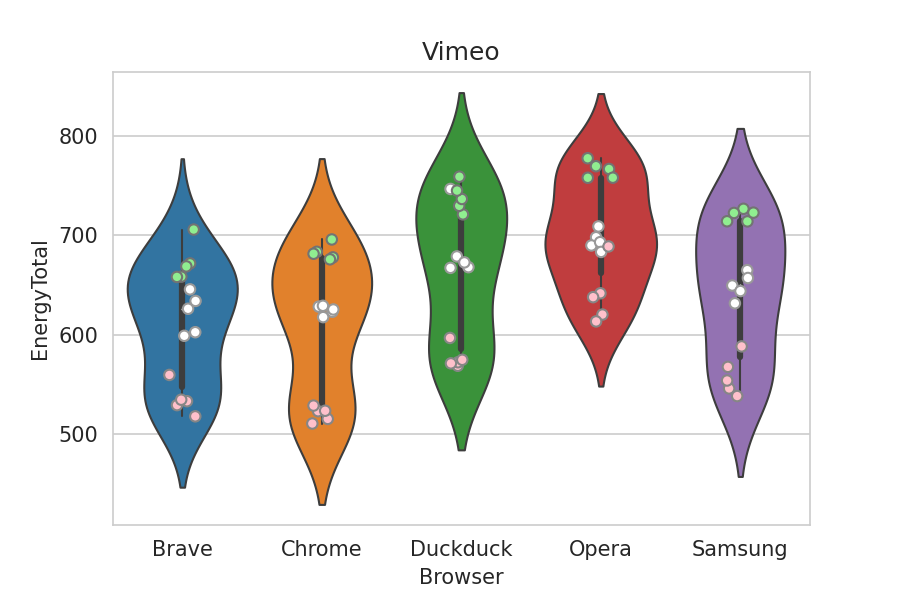}
\caption{Vimeo}
\label{fig:vimeo2}
\end{subfigure}
\begin{subfigure}{0.5\textwidth}
\includegraphics[width=\textwidth]{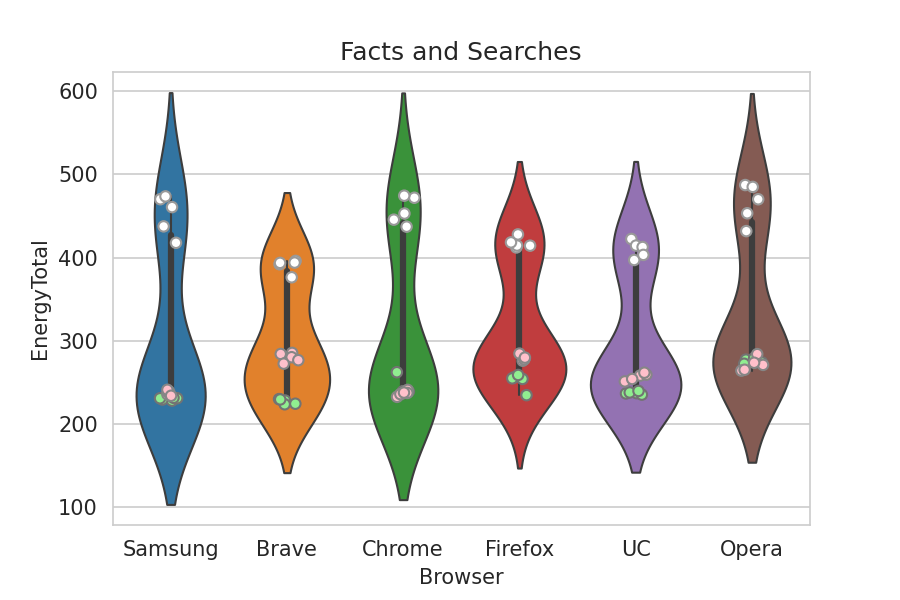}
\caption{Search Engine}
\label{fig:facts2}
\end{subfigure}
\begin{subfigure}{0.5\textwidth}
\includegraphics[width=\textwidth]{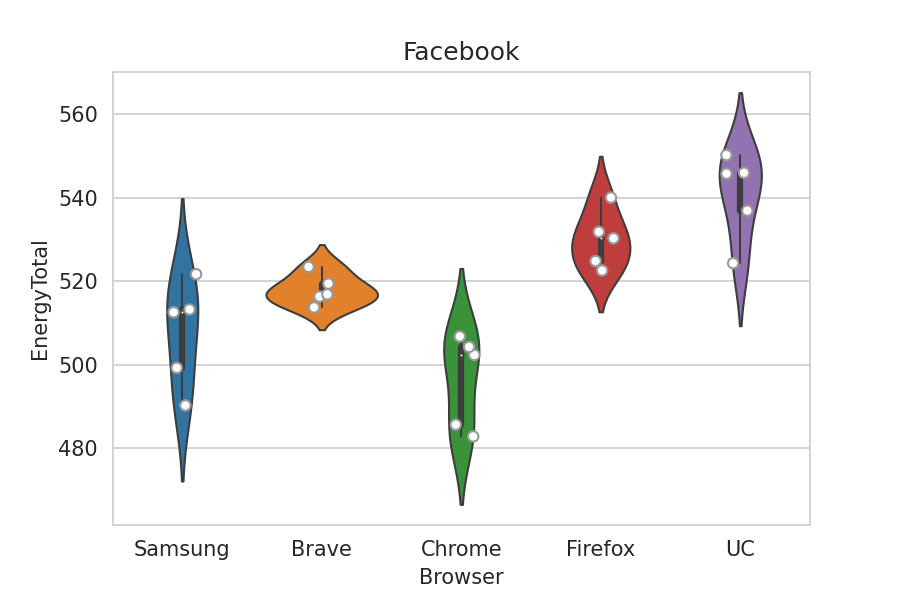}
\caption{Social Media}
\label{fig:facebook2}
\end{subfigure}
\caption{Total energy consumption (Joules) for the different scenarios.}
\label{fig:violin}
\end{figure*}

Figure~\ref{fig:violin} presents violin plots for each browser considered for each scenario. 
These also show the five results of the executions of each case of study in dots. The five dots of the same color represent the five executions of the same script/scenario. The different colors represent the three different scripts/scenarios. With these plots,
it is possible to display the collected data for energy consumption density. Each plot allows understanding the consistency of the results for each browser is consistent through the different usage scenarios. The value of dots
corresponds to the y-axis (Total Energy (Joules)). The chart allows us
to see that the results of the tests performed are very consistent
because the dots of the same color are quite close to each other, although the tests were replicated only five times.

\subsection{Analysis}

In this section we analyze the results from our experiments.

\paragraph{Youtube}
Looking at
Figure~\ref{fig:Youtube}, we can see a discrepancy between the results
obtained. We can see that all executions run in $1144$ seconds,
so the execution time is not a factor that could
influence the results obtained. Firefox was 
the browser that consumed the least energy, $2110$ Joules, while Brave
consumed $2424$ Joules. The difference between Chrome and UC Browser is
almost none, $2266$ and $2265$ Joules respectively, with Samsung Explorer
consuming slightly more than these, $2307$ Joules. Even though Chrome
and Samsung sometimes show advertisements (ads) (closed after five
seconds), which could negatively influence the energy
consumption, they managed to get better results than Brave, which has a
built-in ad-blocker, preventing the appearance of advertisement, thus helping to get
less energy consumption. In the tests performed
for Firefox, no ads appeared when loading the videos, which may have
influenced the lower consumption. The UC Browser, on the other hand,
although no ads were shown, has other default add-ons being loaded
during video loading, which consume more power than Firefox.

\paragraph{Vimeo}
The results obtained in the Vimeo test group, Figure~\ref{fig:vimeo}, show that
for the duration of the $1090$ seconds of the test execution, there is a
significant difference between the browser that consumed the most
energy and the one that consumed the least. Opera Mini consumed in
total $2101$ Joules while Chrome consumed only $1828$ Joules. Opera Mini's
consumption is probably due to the browser's goal, which is to ensure high
performance but save on the space it takes up, and it also has
an ad-block like Brave. To get this increased performance, the
browser consumes more energy. Brave managed to have a very low consumption, almost equaling Chrome, $1829$ Joules,
not being the one that consumed less energy for a minimal
difference. DuckDuck, on the other hand, is another one that consumed
the most energy, being a little behind Opera Mini at $2002$ Joules. DuckDuck, known for its privacy policies, which can be one of the
factors why it has a higher power consumption than the others. The
processes that run in the background to ensure the safety of the
user's navigation end up leading to higher consumption. Samsung
Explorer consumed 100 Joules more than Chrome, that is $1929$ Joules,
showing, as for Youtube, consistency in its consumption, in which it
is not one of the least consuming but also not one of the most
consuming. All tests in this group had no ads loading, which allowed
a better comparison of results.

\paragraph{Search engine}
In Figure~\ref{fig:facts}, we can see that the browser that consumed the most
energy was Opera Mini, consuming $1013$ Joules for $536$ seconds. Brave is
again the least consuming browser with $898$ Joules, 
differing from the UC Browser by $7$ Joules, $905$ Joules. The remaining
browsers have a similar consumption, with Samsung Explorer having $920$
Joules, Chrome $937$ Joules, and Firefox $949$ Joules. Since this group is
about loading pages with information and videos specified above, these
pages may contain ads and therefore may influence the energy
consumption results. Although Opera Mini has a built-in ad-blocker, this
was the browser that consumed the most, compared to Brave, which
is another one with ad-blocking capabilities, consuming  the least. In this respect,
Brave makes its claim of minimizing energy consumption true. Another
aspect that we can see is the loading of web pages. As we have seen in
other works already done, they often keep their focus on performance,
i.e., opening the web page in the shortest time possible, thus
forgetting to reconcile energy consumption. One of the cases that we
can observe is the case of Firefox, which in terms of loading videos,
we saw in the Youtube group that it is the most efficient
browser. Still, here in terms of loading pages, it has already become
the one that consumed more, other than Opera Mini.

\paragraph{Social media}
Finally, Figure~\ref{fig:facebook} shows the results for the Facebook group
where it can be seen that Chrome is the browser that consumed the
least energy, with an energy consumption of $496$ Joules in $295$
seconds. UC Browser has a consumption of $541$ Joules, being the browser
with the highest consumption in the group. Together with UC Browser is
Firefox, which has little difference regarding consumption, with $530$
Joules, making it almost the browser with the highest
consumption. Brave and Samsung Explorer have similar consumption,
consuming $518$ and $507$ Joules, respectively. Even though Brave has an
ad-blocker that sometimes appears on the Facebook web page, it cannot
get better results than Chrome and Samsung Explorer. Once again, there
is a test regarding page loading in this group, and Firefox gets high
results compared to the other browsers. Besides page loading, there is
the loading of images. UC Browser and Brave show a higher consumption
derived from that, considering that the different test strands
specified in design have already been dealt with in the other test
groups.

\subsection{Hypothesis Testing}



In order to validate whether the results obtained  have
statistically significance, we performed statistical
analysis over the collected data. As such, we tested the following
hypotheses:

\begin{center}
\begin{gather*}
	H_0 : P(A > B) = 0.5\\
	H_1 : P(A > B) \neq 0.5
\end{gather*}
\end{center}

The null hypothesis, $H_0$, represents 
that when one randomly draws an object from A and B, the
probability of that object being from A is greater than the probability of being
from B is
$50\%$. The probability is different from $50\%$ in
the alternative hypothesis.

\begin{figure*}[!htb]
\centering
\begin{subfigure}{0.30\textwidth}
\includegraphics[width=\textwidth]{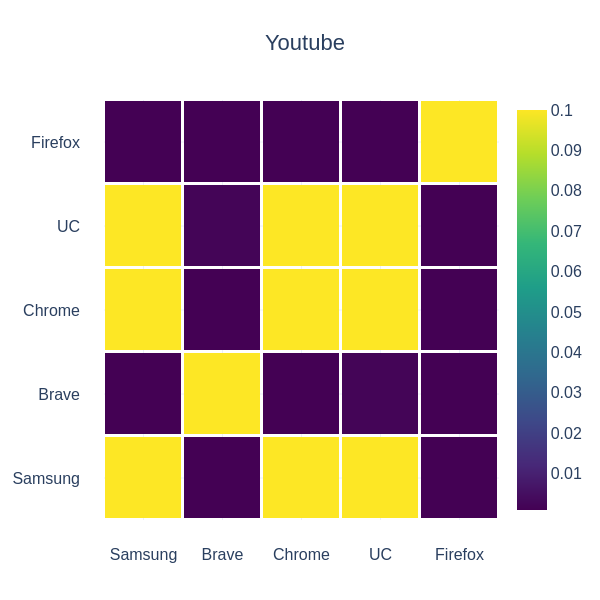}
\caption{Youtube}
\label{fig:Youtube3}
\end{subfigure}
\begin{subfigure}{0.30\textwidth}
\includegraphics[width=\textwidth]{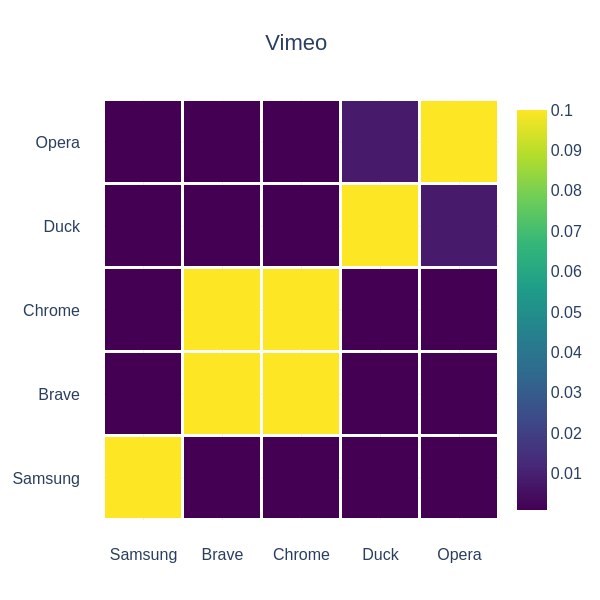}
\caption{Vimeo}
\label{fig:vimeo3}
\end{subfigure}
\begin{subfigure}{0.30\textwidth}
\includegraphics[width=\textwidth]{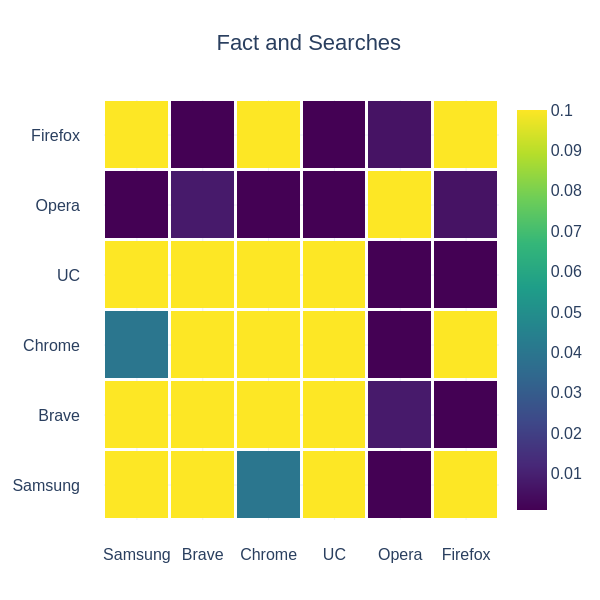}
\caption{Search engine}
\label{fig:facts3}
\end{subfigure}
\caption{Heatmaps with Wilcoxon signed-rank results for the different groups.}
\label{fig:factsxxx}
\end{figure*}

For all the data collected, pairs of browsers were made in the
appropriate test groups in order to understand if there is significant
relevance between the pair difference, i.e., in the Youtube group, we
will get ten pairs; (Samsung, Brave), (Samsung, Chrome), (Samsung, UC),
(Samsung, Firefox), (Brave, Chrome), (Brave, UC), (Brave, Firefox),
(Chrome, UC), (Chrome, Firefox), (UC, Firefox). This same logic
applies to the Vimeo, and search engine groups, with search engine having 15 pairs for six browsers in the test group. These three groups have  independent samples, non-normal distributed and thus we ran the
Wilcoxon signed-rank test with a two-tail \textit{p-value} with $\alpha
=0.05$. The social media group has a different distribution and analyze later. With this in mind, Figure~\ref{fig:factsxxx} presents heatmaps with
the appropriate comparison results for each pair.

In the present heatmaps, it can be seen which pairs of browsers have a
significant difference. Looking at the scale to the right of each
heatmap, the color of the \textit{p-values} $< 0.05$ is indicated, which is the
shades of blue towards purple. In yellow we have the pairs that have
no significant difference. In heatmap of Figure~\ref{fig:Youtube3} one can see that
all pairs have a \textit{p-values} $< 0.01$, except the pairs (Samsung, Chrome),
(Samsung, UC) and (Chrome, UC). In Figure~\ref{fig:vimeo3}, which refers to
Vimeo, the pair (Brave, Chrome) has \textit{p-value} $> 0.05$ and no significant
difference. All other pairs show a highly significant difference. In
Figure~\ref{fig:facts3} the pairs of Opera Mini, (Firefox, Brave) and (Firefox, UC) in
the search engine group, have \textit{p-values} $< 0.01$, while the others
have \textit{p-value} $> 0.05$ and so the difference between them is not
significant. In this group, the differences are not significant,
except when compared with Opera, due to the difference in energy
consumption.

To calculate a non-parametric effect size, Field~\cite{field} suggests
using Rosenthal’s formula~\cite{rose1,rose2} to
compute a correlation, and compare the correlation values against
Cohen’s~\cite{cohen}  thresholds of $0.1$, $0.3$, and $0.5$
for small, medium, and large magnitudes, respectively. For all pairs that
showed a significant difference, with a \textit{p-value} $< 0.05$, we calculated the effect
size and all the results obtained are greater than
$0.5$. We can then conclude that all pairs with a significant
 difference has a large effect size.

In the case of the Facebook group, the data follows a normal
distribution and so it was necessary to use another statistical
method. We used the Analysis of Variance (ANOVA), which is a method that compares the means of more than
2 test groups, that is, in this case, it will serve to compare the
means of the various pairs within the Facebook group. It is thus
possible to analyze whether there is a significant variance between
them. One-Way ANOVA  was the type used for this case because it has
only one independent variable. The \textbf{ANOVA hypotheses} are:

\begin{itemize}
    \item \textbf{Null hypothesis:} Groups means are equal (no variation in means of groups).
    \item \textbf{Alternative hypothesis:} At least, one group mean is different from other groups.
\end{itemize}

\begin{figure}[!htb]
\centering
\includegraphics[width=0.55\columnwidth]{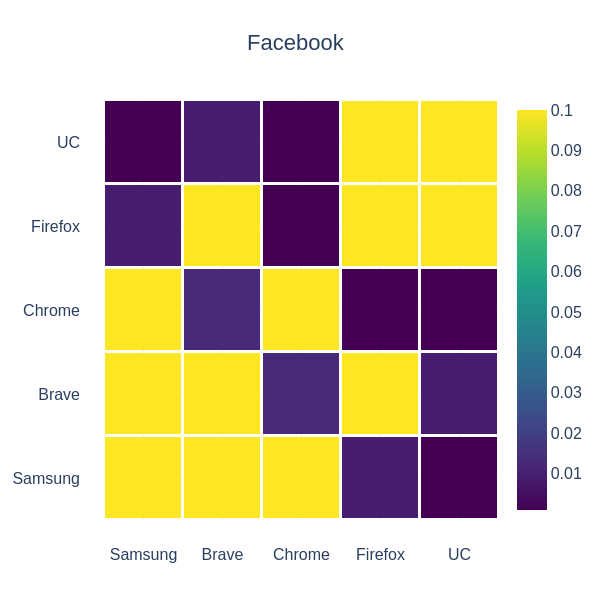}
\caption{Heatmap with ANOVA results for Facebook experiment.}
\label{fig:anova}
\end{figure}

After calculating the \textit{p-value} by using the ANOVA analysis, a \textit{p-value}
of $0.00003$ was obtained. The \textit{p-value} obtained from the ANOVA analysis
is significant ($p < 0.05$), and therefore, we conclude that there are
significant differences among treatments. To know the pairs of
significantly different treatments, we will perform multiple pairwise
comparison (post hoc comparison) analyses for all unplanned
comparisons using Tukey’s honestly significantly differenced (HSD)
test. With this, the \textit{p-values} of each pair were calculated, and these
are represented in Figure~\ref{fig:anova}.

As we can see, the \textit{p-value} calculated through ANOVA analysis,
$0.0003$, is due to most of the compared pairs having a \textit{p-value}
$< 0.05$, except the pairs (Samsung, Brave), (Samsung, Chrome),
(Brave, Firefox) and (Firefox, UC). To understand the effect size of
this analysis, Cohen's F formula was used~\cite{cohen}, where the
basic rules of thumb are that $0.10$ indicates a small effect, $0.25$
indicates a medium effect, and $0.40$ indicates a large effect. The
result of the formula was $0.77$, which means that all pairs with a
significant difference have a large effect size.

\subsection{Research Questions Revisited}

In Table~\ref{tab:browsers}, we can see the summary of browsers'
rankings in each test scenario.

\begin{table}[htb!]
\centering
\caption{Ranking of each browser in each scenario.}
\label{tab:browsers}
\resizebox{\columnwidth}{!}{%
\begin{tabular}{lccccccc}
\toprule
                       & \textbf{Chrome} & \textbf{Samsung} & \textbf{Brave} & \textbf{Firefox} & \textbf{UC} & \textbf{Opera} & \textbf{Duck} \\ \midrule
\textbf{Youtube}       & 3               & 4                & 5              & 1                & 2           &                &               \\ 
\textbf{Vimeo}         & 1               & 3                & 2              &                  &             & 5              & 4             \\ 
\textbf{Search Engine} & 4               & 3                & 1              & 5                & 2           & 6              &               \\ 
\textbf{Facebook}      & 1               & 2                & 3              & 4                & 5           &                &               \\ \midrule
\textbf{Mean}          & 2.25            & 3                & 2.75           & 3.33             & 3           & 5.5            & 4             \\ \bottomrule
\end{tabular}%
}
\end{table}

After analyzing these results, it is possible to answer the five
research questions presented in Section~\ref{sec:intro}.


\paragraph{RQ1: Which mobile browser is the most
  energy-efficient for browsing Youtube?} As we can see, in the
  Youtube test group,  Firefox  was the most
  energy efficient. On Youtube, ads often appear in some browsers,
  such as Chrome and Samsung, influencing their
  consumption. Nevertheless, they had a little different consumption
  from UC Browser, which does not have ads in its executions. Even so,
  Firefox presents a result of reduced energy consumption compared
  with the others. For this task, Brave is the most energy-greedy browser.

 \paragraph{RQ2: Which mobile browser is the most
  energy-efficient for browsing Vimeo?}  For this question, it is
  possible to affirm that Chrome is the most energy-efficient browser. As
  Vimeo never shows ads, Chrome, unlike Youtube, shows the best
  consumption results. Brave and Samsung Explorer are also viable
  options since they differ little from Chrome. Opera Mini occupies the last place in the ranking, being
  the most energy greedy option for browsing Vimeo.

 \paragraph{RQ3: Which mobile browser is the most
  energy-efficient for performing searches in search engines?} In
  this case, we can say that Brave is the most energy-efficient
  browser for searching Google. In this test group, Brave does justice to one of the
  characteristics that it claims to have, consuming less energy than
  the other browsers. UC Browser is also a
  viable option since it presents little difference from Brave. Once
  again, Opera Mini has the highest consumption, being the most energy-greedy alternative
  for the designed tests.

 \paragraph{RQ4: Which mobile browser is the most
  energy-efficient for browsing Facebook?} In this case, it is
  possible to state that the most energy-efficient browser is once
  again Chrome. In this test scenario, there is again the loading of a
  video and, in addition, the loading of images and the news
  feed. Chrome presents a better capacity than the others to manage
  its energy consumption, being the most viable option. Samsung
  Explorer is also a viable option.

 \paragraph{RQ5: Which mobile browser is the most
  energy-efficient overall?} To answer this question, one needs to
  make an overall assessment of our four test scenarios. In Table~\ref{tab:browsers},
  each Browser's rankings were averaged, considering which test
  scenarios they entered, to understand the rankings in total
  better. Although the browsers are not tested in all the test
  scenarios, it is possible to state that Chrome presents the best
  results throughout the four test scenarios. Right after Chorme we have
  Brave, whose results are highly affected by its energy-greedy behavior while browsing Youtube. We conclude that Chrome is the most energy-efficient browser, but Brave is also a viable option, presenting interesting results in several scenarios.


\section{Threats to Validity}
\label{sec:validity}
This work aims to measure the energy consumption of browsers in the
Android environment and compare the results to know which browser is
more energy efficient. We present in this section some threats to the
validity of our study, separated into four categories~\cite{cook}.

\paragraph{Conclusion Validity}
In this category are the threats which may influence the capacity to
draw correct conclusions. It was not possible to compare all the selected
browsers in all groups, and thus there may be cases where a browser not tested
in one group would show better results. Nevertheless, the results
are consistent in almost all the combinations of browsers and tasks, which is also supported by the statistical analysis.

\paragraph{Internal Validity}
This category discusses factors that may interfere with the
results of our study.
A possible issue relates to the measurement of the energy consumed.
We have used a viable tool, Trepn, for
monitoring the execution of the scripts and to get the results. This tool
has been used in many other studies and shown to be reliable \cite{Rua2020,linares,differenttools3}.
While testing, we encountered an initial
problem when executing the scripts. Not all browsers have the same
response time in terms of performance, and therefore, when running the
scripts they had different times for the same steps. To avoid a possible
influence on the results, we
made individual scripts in which the execution time is the same for
all. To execute the scripts we used RERAN, a state-of-the-art tool used in previous studies in the literature, whose event replication has little overhead when compared to real events triggered by actions of real users. The test script was made for each browser in
each test group, giving a total of 52 scripts, each of which was run 5 times, making a total of 260 runs.


\paragraph{Construct Validity}
This section endorses threats that involve generalizing the results to the concept
or theory behind the experiment. Although we cannot generalize the
results for other browsers, we analyzed seven of the most used and relevant browsers.
While it was not possible in the four  test groups to
test all browsers together, the results obtained are consistent.

\paragraph{External Validity}
This category is concerned with the generalization of the results to
industrial practice.
Measurements in different systems
might produce slightly different resulting values if replicated. We
believe these results can be further generalized, and other
researchers can replicate our methodology for future work.

\section{Conclusions}
\label{sec:conc}
This paper presented a preliminary study regarding the energy consumption of web browsers in the Android ecosystem. We studied the energy consumed by seven of the most popular browsers available on Android, by designing and executing a semi-automatic procedure aiming to simulate real user interaction with a browser. For this purpose, we used a record and replay tool of events on the screen to record a set of interactions that intend to simulate the execution of three browser usage scenarios: browsing and viewing videos on streaming video services, performing searches in search engines and browse social networks.

Our preliminary results show that there is not a browser with the best performance for all the tasks considered. We conclude that Google Chrome was the greenest browser, according to a ranking designed for comparing the energy performance of the usage scenarios. Chrome was the most energy-friendly browser for playing videos on Vimeo and browsing Facebook. Surprisingly, Chrome does not
perform well when performing searches in the Google search engine. Nevertheless, Chrome is the greenest browser in our study, with Brave close behind. Finally, we also observed that Opera Mini was the most energy greedy, presenting the worst performance for playing videos on Vimeo and searching on Google.

\section*{Acknowledgment}

This work is ﬁnanced by National Funds through the Portuguese funding
agency, FCT - Fundação para a Ciência e a Tecnologia with references
UIDB/50021/2020 and UIDB/50014/2020, and a PhD scholarship with reference
SFRH/BD/146624/2019.

\clearpage
\balance
\bibliographystyle{IEEEtran}
\bibliography{biblio}

\end{document}